\documentclass[submission,copyright,creativecommons]{eptcs}
\providecommand{\event}{ICLP 2023} 

\usepackage{iftex}

\ifpdf
  \usepackage{underscore}         
  \usepackage[T1]{fontenc}        
\else
  \usepackage{breakurl}           
\fi

\usepackage[space=true]{accsupp}
\usepackage{amsmath}
\usepackage{amssymb}
\usepackage{ dsfont }
\usepackage{graphicx}
\usepackage{hyperref}
\usepackage{ marvosym }
    \usepackage{packages/idpz3syntax}
\usepackage{multirow}
\usepackage{ stmaryrd }
\usepackage{url}
\usepackage{xspace}

\newcommand{\ignore}[1]{}

\newcommand{\marc}[1]{{\color{red} \textbf{Marc:} ``#1''}}
\newcommand{\mvdh}[1]{{\color{mvdhGreen} \textbf{Matthias:} ``#1''}}
\newcommand{\dmar}[1]{{\color{cyan} \textbf{Djordje:} ``#1''}}
\newcommand{\pierre}[1]{{\color{blue} \textbf{Pierre:} ``#1''}}

\newcommand{\voc}{\ensuremath{\Sigma}}
\newcommand{\synrel}{\ensuremath{\sim_s}}

\newcommand{\deref}[1]{\ensuremath{{`#1}}} 

\newcommand{\struct}{\ensuremath{\I}}

\newcommand{\secref}[1]{\textbf{Section}~\ref{#1}}
\newcommand{\FOIL}{\textbf{FOIL}}
\newcommand{\fodot}{FO($\cdot$)\xspace}

\newcommand{\eqcl}[1]{{|#1|_\Ont}}

\newcommand{\intdom}{\ensuremath{\mathcal{C_\Ont}}}
\newcommand{\Ont}{\ensuremath{\mathcal{O}}}
\newcommand{\guards}[1]{\ensuremath{\vdash_{#1}}}

\newcommand{\predi}{\ensuremath{\textsc{pred}}}
\newcommand{\funci}{\ensuremath{\textsc{func}}}
\newcommand{\emptycontext}{\ensuremath{\varnothing}}

\newcommand{\val}[1]{\ensuremath{\$(#1)}}

\newcommand{\ite}[3]{\ensuremath{\mathit{\mathbf{if}}\:#1\:\mathit{\mathbf{then}}\:#2\:\mathit{\mathbf{else}}\:#3}}

\newcommand\m[1]{\ensuremath{#1}\xspace}

\newcommand{\Tr}{\m{\mathbf{t}}}
\newcommand{\Fa}{\m{\mathbf{f}}}

\newcommand{\eval}[1]{\llbracket #1 \rrbracket^{\struct{}}_\nu}

\newcommand{\numeral}[1]{{\mathrm{#1}}}

\newtheorem{definition}{Definition}
\newtheorem{example}{Example}
\newtheorem{theorem}{Theorem}

\title{Quantification and Aggregation over Concepts of the Ontology}
\author{Pierre Carbonnelle \quad Matthias Van der Hallen \quad Marc Denecker
    \institute{KU Leuven\\ Leuven, Belgium\thanks{This research received funding from the Flemish Government under the ``Onderzoeksprogramma Artificiële Intelligentie (AI) Vlaanderen'' programme. The authors thank Maurice Bruynooghe for his review of the drafts of this paper.}}
    \email{pierre.carbonnelle@kuleuven.be \quad matthias.vanderhallen@gmail.com \quad marc.denecker@kuleuven.be}
}
\def\titlerunning{Aggregation over Concepts}
\def\authorrunning{P. Carbonnelle, et al.}

\begin{document}
\maketitle

\begin{abstract}
    We argue that in some KR applications, we want to quantify over sets of {\em concepts} formally represented  by symbols in the vocabulary. We show that this quantification should be distinguished from second-order quantification and meta-programming quantification. We also investigate the relationship with concepts in intensional logic.
    
    We present an extension of first-order logic to support such abstractions, and show that it allows writing expressions of knowledge that are elaboration tolerant.
    To avoid nonsensical sentences in this formalism, we refine the concept of well-formed sentences, and propose a method to verify well-formedness with a complexity that is linear with the number of tokens in the formula.
    
    We have extended \fodot, a Knowledge Representation language, and IDP-Z3, a reasoning engine for \fodot, accordingly.
    We show that this extension was essential in accurately modelling various problem domains in an elaboration-tolerant way, i.e., without reification.
\end{abstract}

\section{Introduction} 

\newcommand{\I}{{\mathcal{I}}}
\newcommand{\fever}{\ensuremath{\mathit{hasFever}}}
\newcommand{\sneezing}{\ensuremath{\mathit{sneezes}}}
\newcommand{\coughing}{\ensuremath{\mathit{coughs}}}
\newcommand{\Symptom}{\ensuremath{\mathit{symptom}}}
\newcommand{\person}{\ensuremath{\mathit{person}}}

The power of a  KR language for compact expression of knowledge lies for an important extent in the way it allows us to abstract over certain types of objects, e.g., to quantify, count, or sum over them.
Classical first-order logic (FO) is a much stronger KR language than Propositional Calculus (PC) because it allows to quantify over domain objects.
Still, the abstraction power of FO is limited, e.g., it does not allow counting or sum over domain objects satisfying some condition.
Many first-order modeling languages are therefore extended to support \emph{aggregations}.
Likewise, one can only quantify over individuals in the domain, not over relations and functions.
This is resolved in second-order logic (SO).

In this paper, we argue that in some KR applications, we want to abstract (i.e., quantify, count, sum, \dots) not over relations and functions, but over certain {\em concepts} in the problem domain, as the following example will show.
\begin{example}
    Consider a corona testing protocol.
    A person is to be tested if she shows at least two of the following symptoms: fever, coughing and sneezing.
    The set of persons showing a symptom is represented by a unary predicate ranging over persons, i.e., $\fever{}/1$, $\coughing{}/1$, and $\sneezing{}/1$.
    Leaving first-order logic behind us, assume that the set of symptoms is represented by the predicate $\Symptom/1$.

    Testing is expressed as a unary predicate $\mathit{test1}/1$, ranging over persons.
    Informally, this predicate is to be defined along the following lines (while freely extending the syntax of FO): 
    \begin{equation}
        \forall p (test1(p) \Leftrightarrow 2\leq  \# \{ x | \Symptom(x) \land x(p)\})
        \label{eq1}
    \end{equation}
\end{example}

The question is: what is the nature of the variable $x$ and the predicate $\Symptom$? Over what sorts of values does $x$ range?

At first sight, one might think that $x$ is a second-order variable ranging over the set of sets of persons, and that we want to count the sets of persons having one of the symptoms and containing person $p$. \ignore{That, at least, was our first impression in this and similar examples. In retrospect, it is not difficult to see that this is \emph{incorrect}.} A `possible world' analysis refutes this idea.
Consider a state of affairs where everybody has all 3 symptoms.
Formally, in a structure $\I$ abstracting such a state, the interpretation of each symptom predicate is identical: it is the set of all persons, $\fever^\I=\coughing^\I=\sneezing^\I=\{y|\person(y)\}$.
Thus, there is only one set that contains person $p$: the cardinality in Eq.~(\ref{eq1}) would be one, and the condition for the first test would not be satisfied for any $p$, even though everyone should be tested.
A different idea is that $x$ is ranging over {\em symbols}, and that we want to count the number of symptom symbols whose interpretation contains person $p$.
This sort of variable is generally found in meta programming, of the kind investigated in, e.g., Logic Programming \cite{BarklundDCL00} or Hilog~\cite{chen1993hilog}.
While this view has merits, it is not totally satisfying.
Assume, for insight, that the team working on this application is international, and the French group has introduced the predicate $\mathit{estFievreux}/1$, French for ``has fever''.
Thus, \fever{}/1 and $\mathit{estFievreux}$/1 are {\em synonymous}.
Consider $\mathit{John}$ whose only symptom is fever.
Both  $\mathit{isFievreux}(\mathit{John})$ and $\fever(\mathit{John})$ are true.
The number of symptom symbols whose interpretation contains $\mathit{John}$ would be two, leading to the erroneous decision that $\mathit{John}$ needs to be tested.

The view that we elaborate in this paper, is that the abstraction in this example 
is over {\em concepts}.
For context, consider that the first phase of a rigorous approach to building a KR specification is the selection of a formal ontology $\voc$  of symbols representing relevant {\em  concepts} of the application field.
These concepts can be identified with the user's \emph{informal interpretations} of the symbols in the ontology of the domain.
The informal interpretation is in practice a crucial concept in all knowledge representation applications: it is the basis of all actions of formal knowledge representation, and of all acts of interpreting formal results of computation in the application domain.
Connecting concepts to informal interpretations provides a good intuitive explanation of what concepts are and how they are relevant in a KR context.




In classical first-order logic, a structure only has the extension of symbols.
The extension of symbol $\sigma$ in a structure is its \emph{formal} interpretation, a.k.a value, in that structure.
It is a relation or function, of a precise arity, over the universe of discourse; in formulae, it is denoted simply by $\sigma$.

For our purpose, we extend a structure to also have the intension of symbols, i.e., an object in the domain of discourse representing its \emph{informal} interpretation.
This intension  is \emph{rigid} in the sense that it is the same in every extended structure.
Since the informal meaning of symbols is known, the synonym relation $\sim_s$ is known too, and it is an equivalence relation  on $\voc$.  Synonymous symbols have the same arity and type.
Synonymous symbols $\sigma_1$, $\sigma_2$ have the same intension, and, in each structure $\I$, the same extension.
In formulae, the intension of $\sigma$ is denoted by $\deref{\sigma}$. 
Finally, we introduce the \emph{value functor} $\$(\cdot)$: when applied to an intension, it returns its formal interpretation in the current structure.

In the example, the intension of the symbols  $\fever/1$ and $\mathit{estFievreux}/1$ are the same object denoted by $\deref{\fever}$ (or $\deref{\mathit{estFievreux}}$) which is the abstraction of the concept of people having fever.
The symbol $\Symptom$ denotes the set containing the intensions  $\deref{\fever}, \deref{\coughing}$ and $\deref{\sneezing}$.
In any particular structure, the $\fever{}/1$ symbol has an interpretation which is a set of elements abstracting the set of people having fever  in the state of affairs.
This set is also the value of the $\deref{\fever}$ intension, i.e., of $\$(\deref{\fever})$.
Suppose we want to count the concepts in $\Symptom$ whose value (a set) contains person $p$.
The following formula expresses the testing protocol:
\begin{equation}
\label{eq:rule1}
    \forall p ( test1(p) \Leftrightarrow 2\leq  \# \{ x | \Symptom(x) \land \$(x)(p)\} )
\end{equation}

In the current state of the art of Knowledge Representation, one might consider, as we once did, applying the common KR technique of \emph{reification}, rather than laboriously extending KR languages as we do.
In this particular case, one could introduce object symbols  $\mathit{Fever}, \mathit{Sneezing}, \mathit{Coughing}$ (under the Unique Name Assumption), and a predicate, $\mathit{Has/2}$.
We can now represent that Bob has fever by the atom $\mathit{Has}(\mathit{Bob},\mathit{Fever})$.
The testing protocols can then be formalized (in a proper extension of FO) as:
\begin{gather}
    \begin{split}
        \forall x ( &\mathit{\Symptom}(x) \Leftrightarrow
        x=\mathit{Fever} \lor x=\mathit{Sneezing} \lor x=\mathit{Coughing})
    \end{split}\\
    \begin{split} \label{eq2}
        \forall p ( \mathit{test1}(x) \Leftrightarrow 2\leq  \# \{ x | \mathit{\Symptom}(x) \land \mathit{Has}(p,x)\} 	)
    \end{split}
\end{gather}

Still, we believe that our study is valuable for two reasons.
First, we improve the scientific understanding of the problem, thus helping avoid using KR encoding tricks to circumvent it.
Second, our logic is more \emph{elaboration tolerant}.
Per \cite{mccarthy1998elaboration}, ``a formalism is elaboration tolerant to the extent that it is convenient to modify a set of facts expressed in the formalism to take into account new phenomena or changed circumstances''.
Because it helps reduce development costs, elaboration tolerance is a desirable feature of formalisms used in AI applications, and in particular in knowledge representation.
\ignore{The simplest kind of elaboration is the addition of new formulas.}
Consider a knowledge base that, initially, does not contain rule~(\ref{eq1}).
Per standard KR practice, it would use predicates such as $\fever{}/1$ to encode symptoms.
Then, circumstances change, and rule~(\ref{eq1}) must be introduced.
A formalism that requires rewriting the knowledge base using reified symptoms $\mathit{Has/2}$ is not elaboration tolerant.
Our formalism, by contrast, would not require any change to the initial knowledge base.

To evaluate our approach, we have extended \fodot to allow quantification and aggregation over concepts, and extended IDP-Z3, a reasoning engine for \fodot, accordingly.  It proved essential to accurately model the knowledge of various problem domains in an elaboration tolerant way, i.e., without reification.  

To summarize, the main contributions of the paper are as follows.
We argue that in some KR applications, it is useful to abstract over concepts in quantification and aggregations.
We analyze the connection to well-known intensional logic (Section~\ref{sec:intensional-logic}), and explain how to distinguish this from second-order quantification and meta-programming (Section~\ref{sec:related}).
We explain how to extend a simple predicate logic so that every ontology symbol has an intensional and an extensional component~(Section~\ref{sec:FoConcept}).
We show that this may lead to nonsensical sentences, and we restrict our syntax to avoid them.
In Section~\ref{sec:FoInt}, we extend FO($\cdot$), a model-based knowledge representation language, and IDP-Z3, a reasoning engine for \fodot, as we propose, and illustrate the benefits of this extension in different applications~(Section \ref{sec:examples}).

\section{``Concepts'' in Logic} 
\label{sec:intensional-logic}

The notions that we study here, i.e., the notions of a ``concept'' and of its ``value'' in a particular state of affairs, are closely related, if not identical,  to the broadly studied distinction between \emph{intension} and \emph{extension}, \emph{meaning} and \emph{designation}, or \emph{sense} and \emph{reference} as studied in philosophical and intensional logic.
The close relation between the two urges us to detail the role of concepts or ``intensions'' in intensional logic, and to make the comparison with our approach.

In philosophical logic, the concepts of \emph{intension} and \emph{extension} have been investigated at least since Frege (for a historical overview, see \cite{sep-logic-intensional}).
A prototypical example concerns the ``morning star'' and ``evening star'', i.e., the star visible in the east around sunrise, and in the west around sunset.
These words have two different intensions (i.e., they refer to two different concepts), but as it happens, in the current state of affairs, their extensions (i.e., their interpretation) are the same: the planet also known as Venus.

Early key contributors in this study are Frege, Church, Carnap and Montague.
The intensional logics of Montague, of Tich\'y, and of Gallin, are typed modal logics based on \emph{Kripke semantics} in which expressions at any level of the type hierarchy have associated intensions.
They provide notations to access the intension and the extension of expressions.
They also provide modal operators to talk about the different values that expressions may have in the \emph{current} and in \emph{accessible} worlds.

The principles of intensions of higher-order objects are similar to those of the base level (i.e., domain elements), leading Fitting in several papers \cite{fitting2004first,sep-logic-intensional} to develop  a simplified logic, called FOIL, where only expressions at the base level of the type hierarchy have intensions.

In intensional logic, the intension of an expression is modelled as a mapping from possible worlds to the interpretation of the expression.
For example, the intension of  $\mathit{EveningStar}$ is the mapping from possible worlds to the extension of $\mathit{EveningStar}$ in that world.
We can express that the \emph{extension} of $\mathit{MorningStar}$ in the current world, equals the \emph{extension} of $\mathit{EveningStar}$ in a (possibly different) \emph{accessible} world:
\begin{equation}
    [\lambda x(\diamond (\mathit{EveningStar}=x))] (\mathit{MorningStar})
\end{equation}
Here the lambda expression binds variable $x$ to the extension of $\mathit{MorningStar}$ in the current world, and the ``$\diamond$'' modal operator is used to indicate an accessible world.
This statement would be true, e.g., in the view of a scientist that may not have evidence that both are the same but \emph{accepts it as a possibility}.

A comparison between the work presented in this paper and the  intensional logics in philosophical logic, is not easy. Even if the notion of intension versus extension is underlying the problems that we study here, the focus in our study is on very different aspects of intensions than in philosophical logic. As  a result, the logic developed here differs strongly from intensional logics.

An important difference is the lack of modal logic machinery in our logic to analyze the difference between intensions and extensions.
In our logic, a symbol has an extension (a.k.a interpretation) in different possible worlds, but there are no modal operators to ``talk'' about the extension in other worlds than the current one.

Another important difference is that we provide an abstraction mechanism through which we can quantify over, and count, intensional objects (a.k.a concepts).
In KR applications, problems such as the symptom example  easily occur (see Section~\ref{sec:examples}).
They can  be analyzed and demonstrated and solved in a  logic much closer to standard logic.

Compared to FOIL, we associate intensions to predicate and function symbols of any arity even though, as hinted earlier, it poses additional challenges to ensure the syntactical correctness of formulae.
We will address these by introducing guards.
By contrast, FOIL associates intensions only to intensional objects of arity 0, like $\mathit{Morningstar}$, avoiding the issue. Finally, our research also includes the extension of an existing  reasoning engine for the proposed logic, capable of reasoning with aggregates over concepts such as in Eq.(3).

\section{FO(Concept)}
\label{sec:FoConcept}

Below, we describe how first-order logic, in its simplest form, can be extended to support quantification over concepts.
The language, FO(Concept), is purposely simple, to explain the essence of our ideas.
We present the syntax~(\ref{sec:syntax}) of the extension, its semantics~(\ref{sec:semantics}), and discuss its complexity and some alternative formulations~(\ref{sec:alt}).


\subsection{Syntax}
\label{sec:syntax}

A vocabulary \voc{} is a set of symbols with associated arity.
We want to extend the FO syntax of terms over vocabulary \voc{} with these four new construction rules:
\begin{itemize}
    \item $\numeral{n}$ is a term if $\numeral{n}$ is a numeral, i.e., a symbol denoting an integer;
    \item $\#\{x_1,\dots,x_n:\phi\}$ is a term if $x_1,\dots,x_n$ are variables and $\phi$ a formula;
    \item $\deref{\sigma}$ is a term if $\sigma \in \voc$,
    \item $\val{x}(t_1,\ldots,t_n)$ is a term if $x$ is a variable, and $t_1, \ldots, t_n$ are terms over \voc.
\end{itemize}

The last rule is problematic, however.
In a structure $\I$ extended as we propose, $x$ ranges not only over the domain of $\I$
, but also over the set of intension of the symbols in \voc{}: when the value of $x$ is such an intension, $\val{x}$ is the extension of the associated symbols; but when $x$ is not such an intension, $\val{x}$ is undefined.
Furthermore, the value assigned to $x$ may be the intension of a predicate symbol: in that case, $\val{x}(t_1,\ldots,t_n)$ is not a term.
Finally, the value of $x$ may be the intension of a function of arity $m \neq n$: in that case, $\val{x}(t_1,\ldots,t_n)$ is not a well-formed term.

For example, $\val{x}()$ is not a well-formed term in the following cases (among others):
\begin{itemize}
    \item[.] $[x=1]$ where $1$ denotes a numeric element of the domain of discourse;
    \item[.] $[x=\deref{p}]$ where $p$ is a predicate symbol;
    \item[.] $[x=\deref{f}]$ where f is a function symbol of arity 1.
\end{itemize}

Essentially, whereas in FO, arities of function and predicate symbols are known from the vocabulary, and taken into account in the definition of well-formed composite terms and atoms by requesting that the number of arguments matches the arity, this now has become impossible due to the lack of information about $\val{x}$.
Thus, to define well-formed terms, we need additional information about the variables occurring in them.
We formalize that information in a typing function.

\newcommand{\proofst}{\guards{t}}
\newcommand{\proofsf}{\guards{f}}

We call $\gamma$ a \emph{typing function} if it maps certain variables $x$ to pairs $(k,n)$ where $k$ is either \predi{} or \funci{}, and $n$ is a natural number.
Informally, when a variable $x$ is mapped to $(k,n)$ by $\gamma$, we know that $x$ is a concept, of kind $k$ and arity $n$.
For a given typing function $\gamma$, we define $\gamma[x:(k,n)]$ to be the partial function $\gamma'$  identical to $\gamma$ except that $\gamma'(x)=(k,n)$.

This allows us to define the notion of well-formed term and formula:

\begin{definition}[Well-formed term]
    We define that a string  $e$  is a \emph{well-formed term} over $\voc$ given a typing function $\gamma$ (denoted $\gamma \proofst e$)  by induction:
\end{definition}
    \begin{itemize}
        \item[.] $\gamma \proofst x$ if $x$ is a variable;
        \item[.] $\gamma \proofst f(t_1,..,t_n)$ if $f$ is an n-ary function symbol of $\voc$ and for each $i$, $\gamma\proofst t_i$;
        \item[+] $\gamma \proofst \numeral{n}$ if $\numeral{n}$ is a numeral, a symbol denoting the corresponding natural number $n$;
        \item[+] $\gamma \proofst \#\{x_1,\dots,x_n:\phi\}$ if $x_1,\dots,x_n$ are variables and $\gamma\proofsf \phi$;
        \item[+] $\gamma \proofst \deref{\sigma}$ if $\sigma \in \voc$;
        \item[+] $\gamma \proofst \$(x)(t_1,\dots,t_n)$ if $x$ is a variable, $\gamma(x)=(\funci{}, n)$ and for each $i \in [1,n]$, $\gamma\proofst t_i$.
    \end{itemize}

(The rules with a ``+'' bullet are those we add to FO's definition of terms.)


\begin{definition}[Well-formed formula]

    We define that a string $\phi$ is a \emph{well-formed formula} over $\voc$ given $\gamma$ (denoted $\gamma \proofsf \phi$) by induction:

\end{definition}
    \begin{itemize}
        \item[.] $\gamma\proofsf \Tr$, $\gamma\proofsf\Fa$;
        \item[.] $\gamma \proofsf p(t_1,\dots,t_n)$ if $p$ is an n-ary predicate of $\voc$ and
            for each $i \in [1,n]$, $\gamma\proofst t_i$;
        \item[.] $\gamma \proofsf (\lnot \phi)$ if $\gamma \proofsf \phi$;
        \item[.] $\gamma \proofsf (\phi \lor \psi)$ if $\gamma \proofsf \phi$ and $\gamma \proofsf \psi$;
        \item[.] $\gamma \proofsf \exists x: \phi$ if $x$ is a variable and $\gamma \proofsf \phi$;
        \item[.] $\gamma \proofsf t_1=t_2$ if   $\gamma\proofst t_1$ and $\gamma\proofst t_2$;
        \item[+] $\gamma \proofsf t_1 \leq t_2$ if $\gamma\proofst t_1$ and $\gamma\proofst t_2$
            and $t_1$ and $t_2$ are numeric terms or cardinality aggregates;
        \item[+] $\gamma \proofsf (\ite{x::k/n}{\phi}{\psi})$ if $x$ is a variable,
            $k$ is either \predi{} or \funci{}, \\
            $n$ is a natural number, $\gamma[x:(k,n)] \proofsf \phi$ and $\gamma\proofsf \psi$;
        \item[+] $\gamma \proofsf \$(x)(t_1,\dots,t_n)$ if $x$ is a variable,
        $\gamma(x)=(\predi{},n)$ and for each $i \in [1,n]$, $\gamma\proofst t_i$.
    \end{itemize}

Notice that, because of the aggregate term rule, these definitions are mutually recursive.
Also, due to their constructive nature, well-formed terms and formulae can be arbitrarily large, but always finite.

Formulae of the form $
    \phi \land \psi, \phi\Rightarrow\psi, \phi\Leftrightarrow\psi, \forall x : \phi
$
are shorthand for these formulae, and are not further discussed:
\begin{equation*}
    \lnot(\lnot \phi \lor \lnot \psi), \neg\phi\lor\psi, (\phi \land \psi) \lor (\neg\phi \land \neg \psi), \neg\exists x : \neg \phi
\end{equation*}
The other comparison operators, $<, >, \ge$, can be defined similarly.

Let $\emptycontext$ be the typing function with empty domain.
We  say  that $\phi$ is a \emph{well-formed formula} over $\voc$ if $\emptycontext\proofsf \phi$.

\begin{example}
    Here is a well-formed version of the cardinality sub-formula of Equation~\ref{eq:rule1}:
    \begin{align*}
        \# \{ x | &\ite{x::\predi{}/1}{Symptom(x) \land \$(x)(p)}{false}\}
    \end{align*}
\end{example}

We say that $Symptom(x) \land \$(x)(p)$ is \emph{guarded} by the $x::\predi{}/1$ condition, and that the formula is \emph{well-guarded}.

Note that, in this logic, a value $\val{x}$ cannot occur in a formula without immediately being applied to a tuple of arguments.

\subsection{Semantics} 
\label{sec:semantics}

\begin{definition}[Intensional ontology and equivalence class $\eqcl{\sigma}$]
    We define an \emph{intensional ontology} \Ont{} as a pair of a vocabulary $\voc$ and a synonym relation $\synrel$ between symbols of same arity in the vocabulary.
    We denote the equivalence class of symbol $\sigma$ in \Ont{} 
    by $\eqcl{\sigma}$,
    and the set of such equivalence classes (or witnesses) by $\intdom{}$.
\end{definition}

In the philosophical papers about intensional logic, the set of concepts is open.
Here, however, we want to quantify only over the concepts that are interpretations of symbols in \voc{}, not over any concepts.
In essence, we want to quantify over $\intdom{}$.

Hence, to define the semantics of FO(Concept), we extend the notion of a structure to include $\intdom{}$ and an additional mapping from concepts to their value.

\begin{definition}[Total structure]
    A \emph{(total) structure} \struct{} over ontology \Ont{} consists of:
\end{definition}
    \begin{itemize}
        \item[.] an object domain $D$ containing the set of natural numbers $\mathds{N}$,
        \item[.] a (total) mapping from predicate symbols $p/n$ in \voc{} to n-ary relations $p^\I$ over $D \cup \intdom{}$,
        \item[.] a (total) mapping from function symbols $f/n$ in \voc{} to $n$-ary functions $f^\I$ over $D \cup \intdom{}$,
        \item[+] a (total) mapping $\$^\struct{}$ from concepts in \intdom{} to relations and functions over $D \cup \intdom{}$.
    \end{itemize}

$\$^\struct{}$ interprets the value function ``$\$(\cdot)$''.


We now define a notion of \emph{coherency} for structures $\struct{}$, expressing that synonyms have the same value or, more formally, that $\$(\deref{\sigma}) = \sigma$ for any $\sigma$.
\begin{definition}[Coherent structure]
    A total structure $\struct{}$ over \Ont{} is \emph{coherent} iff
\end{definition}
    \begin{itemize}
        \item[+] for every predicate symbol $p \in \voc{}$, $\$^\struct{}{(\eqcl{p})} = p^\I$;
        \item[+] for every function symbol $f \in \voc{}$, $\$^\struct{}(\eqcl{f}) = f^\I$. 
    \end{itemize}

From now on, we consider only coherent structures.

We define a \emph{variable assignment} $\nu$ as a mapping of variables to elements in $D \cup \intdom{}$.
A variable assignment extended so that the mapping of $x$ is $d$ is denoted $\nu[x:d]$.

We introduce the ternary \emph{valuation} function, that maps terms $t$ (resp. formulae $\phi$), structures \struct{} and variable assignment $\nu$ to values $v \in D \cup \intdom$ (resp. to Booleans).

\begin{definition}[Value of a term]
    \label{def:term-value}
        We partially define the value $v$ of well-formed $t$ in (\struct{}, $\nu$) (denoted $\eval{t} = v$) by induction:
\end{definition}
\begin{itemize}
    \item[.] $\eval{x}=\nu(x)$ if $x$ is a variable in the domain of $\nu$; 
    \item[.] $\eval{f(t_1, \ldots, t_n)}=f^\I(\eval{t_1}, \ldots, \eval{t_n})$ if
        $f$ is an $n$-ary function symbol and
        $\eval{t_1}, \ldots, \eval{t_n}$ are defined;
    \item[+] $\eval{\numeral{n}}=n$ if $n$ is the integer denoted by $\numeral{n}$;
    \item[+] $\eval{\#\{x_1,\dots,x_n:\phi\}}=m$ if $\llbracket \phi \rrbracket^\I_{\nu[x1:d1]\dots[xn:dn]}$
         is defined for every $d1,\dots, dn \in D \cup \intdom{}$,
         $m$ is an integer, and
         $\#\{x_1,\dots,x_n:\llbracket \phi \rrbracket^\I_{\nu[x1:d1]\dots[xn:dn]} = \Tr\}=m $
    \item[+] $\eval{\deref{S}}=\eqcl{S}$; 
    \item[+] $\eval{\$(x)(t_1,\ldots, t_n)}=\$^\struct{}(\eval{x})(\eval{t_1}, \ldots, \eval{t_n})$ if
        $\eval{x}$ is a concept in \intdom{} mapped by $\$^\struct{}$ to an $n$-ary function,
        and $\eval{t_1}, \ldots, \eval{t_n}$ are defined;
    \item[+] $\eval{t}$ is undefined in all the other cases.
\end{itemize}

\begin{definition}[Truth value of a formula]
    \label{def:form-value}
        We partially define the truth value $v$ of well-formed $\phi$ in (\struct{}, $\nu$) (denoted $\eval{\phi} = v$) by induction:
    \end{definition}

        \begin{itemize}
            \item[.] $\eval{\Tr} = \Tr$; $\eval{\Fa} = \Fa$;
            \item[.] $\eval{p(t_1, \ldots, t_n)}=p^\I(\eval{t_1}, \ldots, \eval{t_n})$ if
                $p$ is an $n$-ary predicate symbol and
                $\eval{t_1}, \ldots, \eval{t_n}$ are defined;
            \item[.] $\eval{\lnot \phi} = \lnot \eval{\phi}$ if $\eval{\phi}$ is defined;
            \item[.] $\eval{\phi \lor \psi} = \eval{\phi} \lor \eval{\psi}$ if $\eval{\phi}$ and $\eval{\psi}$ are defined;
            \item[.] $\eval{\exists x: \phi} = (\exists d \in D \cup \intdom{}: \llbracket \phi \rrbracket^\I_{\nu[x:d]})$ if
                $\llbracket \phi \rrbracket^\I_{\nu[x:d]}$ is defined for every $d\in D \cup \intdom{}$;
             \item[.] $\eval{t_1 = t_2} = (\eval{t_1}=\eval{t_2})$ if
                and $\eval{t_1}$ and $\eval{t_2}$ are defined;
             \item[.] $\eval{t_1 \le t_2} = (\eval{t_1}\le\eval{t_2})$ if
                and $\eval{t_1}$ and $\eval{t_2}$ are defined and integers;
            \item[+] $\eval{\ite{x::k/n}{\phi}{\psi}} = \eval{\phi}$ if $\eval{x}$ is a concept in \intdom{} mapped by $\$^\struct{}$ to an $n$-ary predicate or function according to $k$,
                and $\eval{\phi}$ is defined;
            \item[+] $\eval{\ite{x::k/n}{\phi}{\psi}} = \eval{\psi}$ if $\eval{x}$ is not a concept in \intdom{} mapped by $\$^\struct{}$ to an $n$-ary predicate or function according to $k$,
                and $\eval{\psi}$ is defined;
            \item[+] $\eval{\$(x)(t_1,\dots,t_n)} = \$^\struct{}(\eval{x})(\eval{t_1}, \ldots, \eval{t_n})$ if
                $\eval{x}$ is a concept in \intdom{} mapped by $\$^\struct{}$ to an $n$-ary predicate, and $\eval{t_1}, \ldots, \eval{t_n}$ are defined;
            \item[+] $\eval{\phi}$ is undefined in all the other cases.

        \end{itemize}

    \begin{theorem} 
        The truth value $\eval{\phi}$ is defined for every well-formed formula $\phi$ over \voc{}, every total structure $\struct{}$ over (\voc{}, \synrel), and every variable assignment $\nu$ that assigns a value to all free variables of $\phi$.
    \end{theorem}

    This can be proven by parallel induction over the definitions of well-formed formulae (and terms) and of their value, applying the properties of $\struct{}$ and $\nu$ when needed.
    Indeed, the conditions in the inductive rules of the valuation function match the conditions in the inductive rules of well-formed formulae (and terms).

    A \emph{sentence} is a well-formed formula without free variables.
    We say a total, coherent structure \struct{} \emph{satisfies} sentence $\phi$ iff $\eval{\phi} = \Tr$ for any $\nu$. This is also denoted $\struct \models \phi$.
    Coherent structures \struct{} that satisfy sentence $\phi$ are called \emph{models} of $\phi$.

    \subsection{Complexity and alternative formulations}
    \label{sec:alt}

    The determination of the well-formedness of a formula (Section~\ref{sec:syntax}) can be performed by backward reasoning, using the appropriate inductive rule at each step based on the syntactic structure of the formula under consideration.
    Each inductive step consists of simple syntactical analysis, and, for formulae having sub-formulae, of determining the well-formedness of each sub-formula; formula $\$(x)(t_1,\dots,t_n)$ requires an additional lookup of $x$ in $\gamma$.
    Notice that, at each step, the complexity of the reasoning does not depend on the size of the domain of discourse.
    Bounding the cost of both the syntactical analysis and of the variable lookup by $\alpha$, the cost $C$ of the computation for any formula $\phi$ having sub-formulae $\phi_i$ is bounded as follows: $C(\phi)  \leq \alpha  + \sum_i C(\phi_i) \leq \alpha \times N(\phi)$, where $N(\phi)$ is the number of tokens in the formula ($N(\phi) = 1 + \sum_i N(\phi_i)$).
    Thus, the complexity is linear with the number of tokens in the formula.


    The complexity of decision and search problems in FO(Concept) is the same as in FO.
    This is because the domain $D$ of structures is extended only with a fixed, constant-sized set $\intdom{}$ of elements (and not over higher-order objects).
    For example, deciding the existence of a model of an FO (resp. FO(Concept)) theory with an input domain $D$ (resp. $D  \cup  \intdom{}$) is an NP problem measured in the size of $D$ (resp. $D  \cup  \intdom{}$).

    FO could be extended to support quantification over concepts in other ways than the one presented above.
    Instead of the $\ite{..}{..}{..}$ construct, we could use $\lor$ (and $\land$) with non-strict evaluation.
    Well-guarded quantifications would then be written as follows:
    \begin{align}
        \exists x( (x::\predi{}/1) \land \$(x)(p))\\
        \forall x( (x::\predi{}/1) \implies \$(x)(p))
    \end{align}

    Also, instead of giving \emph{partial} definitions of values (Def.~\ref{def:term-value} and~\ref{def:form-value}), we could give \emph{total} definitions by assigning an arbitrary value when the term (resp. formula) is undefined.
    We would then show that this arbitrary value is not relevant in well-formed formula.

    \section{Extending \fodot with intensions} 
    \label{sec:FoInt}

    We now discuss how we extended the Knowledge Representation language called \fodot~\cite{DBLP:journals/tocl/DeneckerT08, DBLP:conf/cl/Denecker00} to support quantification over intensions.

    \fodot is first-order logic extended with language constructs to make it more expressive:
    \begin{itemize}
        \item[+] types: the vocabulary may include custom types, in addition to the built-in types (e.g., \lstinline|Int, Bool|).  Each symbol has a type signature of the form
        \lstinline[language=IDP]|T1**...**Tn->T|
        specifying their domain and range.
        The range of a predicate is the set of Booleans \lstinline|Bool|.
         Formula must be well-typed, i.e., predicates and functions must be applied to arguments of the correct type.
        \item[+] equality: \lstinline|t1=t2| is a formula, where \lstinline|t1| and \lstinline|t2| are terms of the same type.
        \item[+] arithmetic over integers and rationals: arithmetic operators (\lstinline|+,-,*,/|) and comparisons (e.g., \lstinline|=<|) are interpreted functions.
        \item[+] binary quantification: \lstinline|?x \in P: p(x).| (where \lstinline|P| is a type or predicate) is equivalent to \lstinline|?x: P(x) & p(x).|
        \item[+] aggregations, such as \lstinline|#{x \in P: p(x)}| (count of \lstinline|x| in \lstinline|P| satisfying \lstinline|p|) and \lstinline|sum{{f(x) \mid x \in P}}| (sum of \lstinline|f(x)| over \lstinline|x| in \lstinline|P|).
        \item[+] (inductive) definitions~\cite{DBLP:journals/tocl/DeneckerT08}: \fodot theory consists of a set of logic sentences and a set of (potentially inductive) definitions. Such a definition is represented as a set of rules of the form: \\
        \lstinline|    ! x1 in T1,..,xn in Tn: p(x1,..,xn) <- F.|\\
        where F is an \fodot formula.
    \end{itemize}

    \ignore{
        In \fodot, identifiers, i.e., nullary symbols denoting the same elements of $D \cup \mathcal{I}$ in all interpretations are not followed by $()$.}

    Thanks to the support of inductive definitions, \fodot is unique in combining the expressivity of classical logic and logic programming~\cite{DBLP:conf/cl/Denecker00}.
    The reference manual of \fodot is available online\footnote{\url{https://fo-dot.readthedocs.io/}}.

    A Knowledge Base (KB) written in \fodot cannot be run: like human knowledge, it is just a “bag of information”, formally describing models in a problem domain.
    Knowledge bases do not distinguish inputs from outputs, and allow reasoning in any direction.
    They are used to perform a variety of reasoning tasks (each with a particular set of inputs and outputs), using generic methods provided by reasoning engines, such as IDP-Z3\footnote{\url{http://idp-z3.be/}} and FOLASP~\cite{DBLP:journals/corr/abs-2108-04020}: they can find relevant questions to solve a particular problem, derive the consequences of new information, explain how they derived these consequences, find possible models, and find a model that minimizes a cost function~\cite{DBLP:books/mc/18/Cat0BJD18}.

    These generic capabilities are used to easily build knowledge-based interactive systems that assist users in finding solutions to problems in a problem domain~\cite{DBLP:conf/ruleml/DassevilleJJVD16,DBLP:conf/semco/DeryckVDM19}.

    To allow reasoning about concepts, we have extended \fodot with the `` \lstinline|`.| '' operator (to refer to the intension of a symbol) and the ``\lstinline|$(.)|'' operator (to refer to the interpretation of a concept), as described in Section 3 for FO.

    An issue arises in expressions of the form
    \begin{lstlisting}[language=IDP]
        $(x)(t1,...,tn)
    \end{lstlisting}
    $x$ must be a \lstinline|Concept|, the arguments \lstinline|t1,...,tn| must be of appropriate types and number for the predicate or function \lstinline|$(x)|, and \lstinline|$(x)(t1,...,tn)| must be of the type expected by its parent expression.

    To address this issue and to support the writing of well-guarded formulae, we have introduced types for the concepts having a particular type signature:
    \begin{lstlisting}[language=IDP]
        Concept[T1**..**Tn -> T]
    \end{lstlisting}
    The interpretation of this type is the set of concepts with signature \lstinline|T1**..**Tn->T|.
    Note that \lstinline|T1,..,Tn,T| themselves can be conceptual types.

    The well-formedness and semantics of quantifications over a conceptual type :
    \begin{lstlisting}[language=IDP]
        ?x\in Concept[T1**..**Tn -> T]: $(x)(t1, .., tn).
    \end{lstlisting}
    is defined by extending the concept of guards (Section~\ref{sec:syntax}), and considering the following equivalent statement with guards:
    \begin{lstlisting}[language=IDP]
        ?x: if x::[T1**..**Tn -> T] then $(x)(t1,..,tn) else false.
    \end{lstlisting}
    where \lstinline|x::[T1**..**Tn -> T]| is a guard specifying the types of the arguments of $x$, and the type of \lstinline|$(x)(t1,...,tn)|.
    The definitions of the typing function $\gamma$ (Section~\ref{sec:syntax}), and of well-formed formulae and their semantics are updated accordingly.

    The syntax of \fodot allows applying the ``\lstinline|$(.)|'' operator to expressions (not just to variables), e.g., in atom \lstinline|$(expr)()|.
    The type of \lstinline|expr| must be a conceptual type with appropriate signature.



    We have updated the IDP-Z3 reasoning engine\footnote{\url{http://idp-z3.be/}} to support such quantification over concepts.
    IDP-Z3 transforms \fodot theories into the input language of the Z3 SMT solver~\cite{de2008z3} to perform various reasoning tasks.
    A quantification
    \begin{lstlisting}[language=IDP]
        ? x in Concept[T1**..**Tn->T]: expr(x).
    \end{lstlisting}
    is transformed into a disjunction of \lstinline|expr(`c)| expressions, where \lstinline|`c| is a \lstinline|Concept[T1**..**Tn->T]|.
    Occurrences of \lstinline|$(`c)(t1,..tn)|
    within \lstinline|expr(`c)| are transformed into \lstinline|c(t1,..tn)| where \lstinline|c| is the symbol denoting concept \lstinline|`c|
    (Our implementation currently does not support synonymous concepts).
    The resulting formula is an FO sentence that can be submitted to Z3.


    \section{Examples in \fodot} 
    \label{sec:examples}

    In our practice, we have identified a few examples where quantifications over concepts proved essential to accurately model the knowledge available within a domain in an elaboration tolerant way, i.e., without reification.
    These examples come from a broad range of applications.
    They are:
    \begin{itemize}
        \item the May 2021 DM Community challenge\footnote{\url{https://dmcommunity.org/challenge/challenge-may-2021/}} about deciding to perform additional testing of patients, based on a set of symptoms (similar to the example we used in the introduction);
        \item the ``International Law to fight money laundering'' example, where we want to represent the general rule that the national laws must be stricter than the EU directive.
        \item the ``Word disambiguation'' example, in which the word ``child'' in a statutory law could represent either the biological or legal child.
        \item the ``template'' example, in which intensional objects are used to define templates.
    \end{itemize}
    The examples are available online\footnote{\url{https://tinyurl.com/Intensions}}, and can be run using the IDP-Z3 reasoning engine.
    Below, we discuss the International law example for illustration purposes.

        Since 1990, the European Union has adopted legislation to fight against money laundering and terrorist financing.
        It creates various obligations for the parties in a business relationship, such as verifying the identity of the counter-party.
        The member states have to transpose the directive into national laws.
        The national laws must meet the minimum obligations set forth in the EU directive.

        In our simplified example, the EU directive requires the verification of identity in any transaction with a value above 1M\EURdig;  a national law might set the threshold at 500K\EURdig instead.
        Similarly, the EU directive might require a bank to send a report to their authority at least quarterly, but a country might require a monthly report.

        Our goal is then to express the requirement that the national obligations are stricter than the EU ones.
        We choose an ontology in which ``has a lower value'' means ``stricter''.
        We also use a mapping from the parameters of the national laws to their equivalent parameter in the EU law.

        \begin{lstlisting}[language=IDP]
        vocabulary {
            type Country
            threshold, period: Country -> Int
            obligation: Concept[Country->Int]->Bool
            thresholdEU, periodEU: () -> Int
            mapping: Concept[Country->Int]-> Concept[()->Int]
        }
        theory {
            obligation := {`threshold, `period}
            mapping := {`threshold ->`thresholdEU, `period->`periodEU}

            // national law must be stricter than European law.
            !o\inobligation: !c\inCountry: $(o)(c)=<$(mapping(o))().
        }
        \end{lstlisting}


        Notice that an expression (not a variable) is applied to the value operator: \lstinline|$(mapping(o))()|.
        Because \lstinline|o| is a \lstinline|Concept[Country->Int]| by quantification, it is in the domain of \lstinline|mapping|, and \lstinline|mapping(o)| has type \lstinline|Concept[()->Int]|.
        Its value is thus a nullary function of range \lstinline|Int|.

    \section{Related Work} 
    \label{sec:related}

    We discussed the relation between intensional logic and our work in Section~\ref{sec:intensional-logic}.
    We now discuss the relation with other work.

    \subsection{Second-order quantification and relations}

    A contribution of this work is to clarify the utility of being able to quantify over concepts in the vocabulary and to show how it differs from quantifying over sets or functions as in second-order quantification.
    Certainly in the case of predicate or function intensions, it is our own experience that it is easy to confuse the two.
    We showed in the introduction the need for quantification over intensions as opposed to over second or higher-order objects (relations or functions).
    However, the quantification over intensions cannot replace quantification over second-order.
    A clear-cut example of second-order quantification and relation occurs in the \emph{graph mining problem} \cite{DBLP:journals/amai/HallenPJD19}.

    \begin{example}
    A graph \emph{homomorphism} $h$ is a function from the nodes of a graph, called the \emph{pattern} $p$, to nodes of a graph $g$, that preserves the edges.
    The following expression intends to define $\mathit{hom/1}$ as the set of homomorphisms of $p$ in $g$, where $p(x,y)$ (resp. $g(x,y)$) denotes an edge between nodes $x$ and $y$.
        \[\forall h( hom(h) \Leftrightarrow \forall x\forall y( p(x,y) \Leftrightarrow g(h(x),h(y))))\]
    \end{example}

    The question is: over what sorts of values does $h$ range (in the context of a structure $\I$)?  Over the set of  intensions of unary function symbols in $\voc$? There may be none! Or over the set of all binary functions from nodes of the pattern to nodes of the graph in $\I$?  Clearly, over the latter. Hence, $h$ is a second-order variable and $\mathit{hom/1}$ a second-order predicate symbol.

    \subsection{Metaprogramming}

    Variables ranging over concepts, as we propose, are similar to variables ranging over symbols, as found in meta programming.
    Indeed, meta programming also brings the benefit that we seek: formulating knowledge about concepts in an elaboration tolerant way (i.e, without reification).

    Meta programming has been proposed in logic-based languages such as OWL \cite{DBLP:journals/logcom/Motik07,DBLP:conf/owled/PanHS05}, as well as in rule-based languages such as Prolog~\cite{BarklundDCL00}, HiLog~\cite{chen1993hilog}, or webdamlog~\cite{DBLP:conf/pods/AbiteboulBGA11}, but not in languages integrating the two types of formulation, such as \fodot.
    This integration is needed to support all the use cases that we describe.
    For example, the International Law example encodes a logic constraint that cannot be encoded in a rule-based system; the Template example encodes an inductive definition that cannot be encoded in logic-based systems.
    By contrast, our reasoning engine does support the two types of formulation, and all the use cases.

    Furthermore, we formalize the guarding mechanism that ensures that sentences are well-formed.
    To the best of our knowledge, this mechanism is missing in all meta-programming implementations (and in particular in HiLog and webdamlog).
    Without this mechanism, the developer of a knowledge base cannot benefit from the automatic detection of syntactical errors.

    \section{Summary} 
    As our examples illustrate, it is often useful in knowledge-intensive applications to quantify over concepts, i.e., over the intensions of the symbols in an ontology.
    An intension is an atomic object representing the informal interpretation of the symbol in the application domain.

    First-order logic and \fodot can be extended to allow such quantifications in an elaboration tolerant way.
    Appropriate guards should be used to ensure that formulae in such extensions are well-formed: we propose a method to verify such well-formedness with a complexity that is linear with the number of tokens in the formula.


    While related to modal logic, the logic introduced here differs strongly from it.
    It has no modal operators to talk about the extension of a symbol in other worlds than the contextual one, but it offers mechanisms to quantify and count intensional objects.

\nocite{*}
\bibliographystyle{eptcs}
\bibliography{references.bib}
\end{document}